\documentclass[twocolumn,letterpaper]{IEEEAerospaceCLS}  

\usepackage[]{graphicx}    
\newcommand{\ignore}[1]{}  
\usepackage{times}
\usepackage{url}
\usepackage[hidelinks]{hyperref}
\usepackage[utf8]{inputenc}
\usepackage[small]{caption}
\usepackage{graphicx}
\usepackage{amsmath}
\usepackage{tikz}
\usepackage{pgfplots}
\usepackage[detect-all,per-mode=symbol,binary-units=true]{siunitx}
\usepackage{algorithm}
\usepackage[noend]{algpseudocode}
\usepackage{multirow}
\usepackage{cleveref}
\usepackage{vector}
\usepackage{placeins}
\usepackage{booktabs}
\usepackage{ctable}
\usepackage{tablefootnote}

\usetikzlibrary{automata,shapes,arrows}
\tikzset{mnode/.style={circle, draw=black, minimum size=0.5cm}}

\begin{document}
\title{Markov Decision Processes For \\ Multi-Objective Satellite Task Planning}

\author{%
Duncan Eddy\\ 
Stanford University\\
496 Lomita Mall\\
Durand Building\\
Stanford, CA 94305\\
deddy@stanford.edu
\and 
Mykel Kochenderfer\\
Stanford University\\
496 Lomita Mall\\
Durand Building\\
Stanford, CA 94305\\
mykel@stanford.edu
}

\maketitle

\thispagestyle{plain}
\pagestyle{plain}

\setcounter{footnote}{0}

\begin{abstract}
This paper presents a semi-Markov decision process (SMDP) formulation of the satellite task scheduling problem. This formulation can consider multiple operational objectives simultaneously and plan transitions between distinct functional modes. We consider the problem of scheduling image collections, ground contacts, sun-pointed periods for battery recharging, and data recorder management for an agile, resource-constrained Earth-observing spacecraft. By considering multiple mission objectives simultaneously, the algorithm is able to find optimized task schedule that satisfies all operational constraints in a single planning step, thus reducing the operational complexity and number of steps involved in mission planning. We present two solution approaches based on forward search and Monte Carlo Tree search. We baseline against rule-baed, graph search, and mixed-integer-linear programming approaches. The SMDP formulation is evaluated in both single-objective and multi-objective scenarios. The SMDP solution is found to perform comparably with the baseline methods at greatly increased speed in single-objective scenarios and greater schedule reward in multi-objective.


\end{abstract}

\tableofcontents

\section{Introduction}

In recent years, small satellites have demonstrated their value as Earth observation platforms capable of hosting a variety of metrology payloads. Dividing the workload of a single large satellite between multiple smaller satellites has been shown to reduce cost, provide robustness to single-point-failures, decrease revisit times, and increase information gathering throughput \cite{brown2006value,derrico2012distributed}. These small satellite platforms typically operate in low Earth orbit (LEO). Operating in this orbit regime presents its own unique set of challenges. Since direct contact with a satellite is limited to brief periods over ground stations, longer term task schedules must be generated for the vehicle to continuously operate.
This paper studies the problem of generating task schedules using a semi-Markov decision process formulation.

Satellite task scheduling involves choosing data collection opportunities to maximize an observation objective given a set of target locations to observe, the satellite trajectory, and a planning horizon. Common objectives include maximizing the number of images collected, timeliness of data return, or total monetary reward. There are often constraints on ground or on-board resources. These constraints are particularly relevant to small satellites as the reduction in platform size is associated with reductions in power generation, energy storage, and on-board data storage capacity. Time constrained task-sequencing has been shown to be NP-complete \cite{garey1979computers}; consequently, various formulations and heuristics have been introduced over the years \cite{hall1994maximizing,lemaitre2002selecting,bianchessi2005earth,beaumet2011feasibility}.

Past work has largely focused on rule-based \cite{bensana1999dealing,bianchessi2008planning} or integer programming \cite{augenstein2016optimal,nag2018scheduling,cho2018optimization,shah2019scheduling} formulations of the satellite task planning problem. Rule-based systems are advantageous in that they are computationally efficient, are straight forward to build, and can guarantee specific behaviors. However, rule-based systems and heuristics suffer from the fact that it is not obvious how to formulate the rules to achieve optimal system performance and such systems cannot dynamically trade between multiple disparate operational objectives. 
All decision permutations have to be explicitly considered by the designer, which can lead to complex logic to encode operational rules.

Mixed-Integer Linear Programming (MILP) approaches guarantee that a solution is optimal up to the dual bound under the modeling assumptions used to formulate the problem. The guarantee that the resulting plan is best, along with the ability to enforce specific  constraints, makes this approach attractive for many missions. The draw-backs  are that it is computationally intensive and does not efficiently scale with the number of satellites or the number of decision variables. Furthermore, it is an inherently discrete-time model of the system and must rely on bounding-value approximations for power and data resource usage. 

 More recent work has started to apply deep reinforcement learning techniques to address challenges related to satellite task scheduling. Reinforcement learning (RL) is a field of machine learning where an agent learns a decision policy through repeated interactions with an environment. The problem is typically modeled as a Markov Decision Process (MDP) \cite{sutton2018reinforcement}, where reasoning is done in terms of the agent's transition and reward functions.
  
 Ferreira et al. \cite{ferreira2018multiobjective} applied reinforcement learning to manage the optimization of selecting radio control parameters for a satellite communications system in various operational scenarios. In the Earth Observation (EO) domain, Hadj-Salah et al. \cite{hadjsalah2019schedule} applied reinforcement learning to optimize image collection planning in the presence of cloud coverage uncertainty.
 
 
 This paper studies a model-based approach with a known reward function. The problem is posed as a semi-Markov Decision Process (SMDP). An SMDP is a variant of an MDP that reduces computational complexity by only evaluating the decision policy at specific times. We introduce two different approximate solution techniques to the problem\textemdash forward search and Monte Carlo Tree Search (MCTS). The viability of this approach is demonstrated by comparison to common single-objective optimization approaches, and results are presented for multi-objective optimization as well.
 
\section{Markov Decision Process Formulation}

A Markov Decision Process is a general framework \cite{sutton2018reinforcement,kochenderfer2015decision} for modeling and solving decision making problems. The agent (the satellite) chooses an action in the current state, receives a reward, and then transitions to the next state. The process is repeated over the planning horizon. The state space $\mathcal{S}$, action space $\mathcal{A}$, transition function $T$, and reward function $R$, define an MDP 
\begin{equation}
	\mathcal{M} = (\mathcal{S},\mathcal{A},\mathit{T},\mathit{R})
\label{eqn:mdp_problem}
\end{equation}

Relevant to the Earth observing satellite tasking problem is a set of \emph{locations} $I$, which have been requested to be collected over a time interval $[0, H]$ with $H$ as the planning horizon. Each image $i \in I$ is defined by an Earth-fixed center point and has an associated reward for collection $r_i$. Over the horizon, each image has a set of windows of \emph{opportunity} $O_i$. Each individual opportunity for image $i$ is denoted $o_i \in O_i$ and has a start time $t_s$ and end time $t_e$ during which the image could be collected. The problem is to decide which images should be collected to maximize the total reward over the planning horizon.

There are a few notable variations of this core problem. The first is the agile satellite scheduling problem. Here, the satellite imaging instrument is fixed in the satellite body-frame and the satellite must use its attitude control system to reorient to point at each location before collecting an image. The second variation adds the complication of ground contact planning; the satellite can only be communicated with while it is in view of a set of ground station locations $G$, and ground contacts must be scheduled to download collected images. The final variation considered here is the imposition of limitations on on-board spacecraft resources, in particular constraints on power and data.

One challenge of modeling the satellite tasking problem as an MDP is that typical planning horizons are long, while the required time step for making decisions is significantly smaller. Planning horizons are, at a minimum, a few hours long\textemdash the average time between ground contacts in Low Earth Orbit. More useful scheduling horizons are on the order of days or weeks. However, opportunities to collect an image or make contact with the ground occur at specific times, when the decision to take an opportunity must be made. Since the duration of these opportunities are minutes or seconds long, the decision making time step has to have the same or finer resolution. The standard MDP formulation therefore has an extremely large state space, making it computationally challenging to solve. The standard formulation results in a difficult exploration problem, where the rewards associated with collecting or downlinking images are greatly delayed compared to the current agent state. 

\begin{figure}
\begin{center}
	\begin{tikzpicture}
    \draw[->] (0,1) |- (0.5,1) node(xline)[right] {$t$};
    \draw[] (0,0.9) -- (0,1.1) node(xline) {};

    \node[mnode] at (0, 0)  (a) {$t_0$};
    \node[mnode] at (1, 0)  (b) {$t_1$};
    \node[mnode] at (2, 0)  (c) {$t_2$};
    \node[mnode] at (3, 0)  (d) {$t_3$};
    \node[mnode] at (4, 0)  (e) {$t_4$};

    \draw[->] (a) -- (b);
    \draw[->] (b) -- (c);
    \draw[->] (c) -- (d);
    \draw[->] (d) -- (e);

    \node[mnode] at (0, -1.5)  (A) {$t_0$};
    \node[mnode] at (2.7, -1.5)  (B) {$t_1$};
    \node[mnode] at (4.62, -1.5)  (C) {$t_2$};

    \draw[->] (A) edge [bend left] node[right] {} (B);
    \draw[->] (B) -- (C);
    \draw[->] (A) edge [bend right] node[right] {} (C);
\end{tikzpicture}	
\end{center}
\caption{Illustration of the difference of time discretization in MDP and SMDP formulations. In the MDP formulation (top), all time steps are uniform and are taken sequentially. In the SMDP model (bottom) transitions occur at non-uniform times and some time steps are skipped.}
\label{fig:mdp_smdp}
\end{figure}
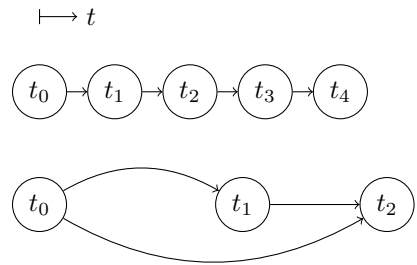

To address the challenge of computational tractability we take advantage of the inherent structure of the problem where opportunities for image collection or communication with the ground only occur at specific times. This property indicates that the problem could be posed as a semi-Markov Decision Process, a variation of the MDP. In a semi-Markov model, the uniform time step assumption inherent to MDPs is removed, and the agent state is instead only considered at times when actions could be taken. See \Cref{fig:mdp_smdp} illustrates this difference. This modeling choice reduces the size of the decision space, making the problem more computationally tractable, and also removes the challenge of delayed rewards. The rest of the section describes the state, action, transition, and rewards of the SMDP formulation, as well as the solution approach.

\subsection{State Space}

The satellite state is defined as
\begin{equation}
s = (t, t_s^p, I_{c}, d, p)
\label{eqn:state_def}
\end{equation}
where $t$ is the time of the most recent action taken, $t_s^p$ is the start time of the last collection or downlink opportunity taken, $I_{c}$ is the set of all images collected by time $t$, $p$ is current spacecraft power, and $d$ is the current on-board data usage. Data and power can be removed from the state definition if resources are not included as part of the problem.

\subsection{Action Space}

At each discrete step $t$, the agent selects the next action to take from the set of feasible actions for that state $A(s)$, where $A(s) \subset \mathcal{A}$. For a semi-Markov model the next time step directly corresponds to the start time $t_s$ of the next action. The possible actions are drawn from an enumeration of possible higher-level functional modes the satellite can assume. For this problem we consider actions of (image) collect, (ground) contact, and sun-point. This action set can be extended to include other common modes like maneuvering.

Actions are defined as
\begin{equation}
a = (t_s, t_e, \ell)
\label{eqn:action_def}
\end{equation}
where $t_s$ is the start-time, $t_e$ is the action end-time, and $\ell$ is the location. The location $\ell \in \left\{i, g, \mathsc{nil}  \right\}$ indicates which image $i \in I$ or ground station $g \in G$ the action is associated with. For sun-point actions the location is $\mathsc{nil}$.  This information is also used to indicate the mode associated with the action. It is assumed that for all collect and contact actions there is also a sun-point action with the same start and end times that is always available to take. This might not always be be useful action to take (e.g. if the satellite is in eclipse) but it is still an option. 

For each state, the action space $A(s)$ is the set of all possible actions with start-time after the current time that are feasible for the spacecraft to transition to given the last collection or contact time $t_s^p$:
\begin{equation}
A(s) = \left\{a \; | \; t_s > t, C(t_s^p, t_s) = 1 \right\}
\label{eqn:action_space}
\end{equation}
$C(t_s^p, t_s)$ is the agility constraint function that indicates whether a slew between the pointing configurations for actions with start times $t_s^p$ and $t_s$ is feasible. It is a binary function $C(t_s^p, t_s) \in \left\{0, 1\right\}$ where $1$ denotes feasibility and $0$ infeasibility of the transition.

\subsection{Transition Function}

$T: \mathcal{S} \times \mathcal{A} \rightarrow \mathcal{S}$ is the transition function used in this work. Given an input state and action, the transition deterministically returns the next state. For the semi-Markov formulation, this means advancing the agent state to $t_s$, the start time of the chosen action\footnote{The end time is equally valid from an implementation point-of-view so long as the times are consistent throughout implementation.}. The transition also updates on-board resource states by propagating their dynamics. For this work, we use a simple linear update for power and data resources. \Cref{eqn:time_update,eqn:past_action_update,eqn:collected_update,eqn:collected_update,eqn:power_update,eqn:data_update} provide the transition that propagates the state $s_t$ to the next state $s_{t+1}$ after taking action $a$:
\begin{equation}
t \gets t_s
\label{eqn:time_update}
\end{equation}
\begin{equation}
t_s^p \gets
\begin{cases} 
      t & \text{if $\ell \in I$ or $\ell \in G$} \\
      t_s^p & \text{otherwise}
   \end{cases}
\label{eqn:past_action_update}
\end{equation}
\begin{equation}
I_c \gets
\begin{cases} 
      I_c \cup \{\ell\} & \text{if $\ell \in I$ and $\ell \notin I_c$} \\
      \phantom{-} & \text{and $p > p_{\text{min}}$} \\
      \phantom{-} & \text{and $d < d_{\text{max}}$} \\
      I_c & \text{otherwise}
   \end{cases}
\label{eqn:collected_update}
\end{equation}
\begin{equation}
p \gets p + (t_s - t_s^p)\dot{p}_\ell
\label{eqn:power_update}
\end{equation}
\begin{equation}
d \gets d + (t_s - t_s^p)\dot{d}_\ell
\label{eqn:data_update}
\end{equation}

\Cref{eqn:time_update} advances the time. \Cref{eqn:past_action_update} updates the most recent collect or contact action if a collect or contact was taken. \Cref{eqn:collected_update} adds an image to the set of collected images if it had not been previously collected and there are enough available resources to do so. \Cref{eqn:power_update,eqn:data_update} could also be exchanged for non-linear update functions that capture full system dynamics to the desired degree of fidelity. Here, $\dot{p}_\ell$ and $\dot{d}_\ell$ are the power and data generation associated with the action type. The exact values of each depend on the satellite design, but the general characteristics of each mode are summarized in \Cref{tab:resource_states}. During image collections power is consumed and data generated, during ground contacts power is again consumed but data is removed, and finally during sun-point actions power is generated along with data from the continuous recording of spacecraft telemetry.

\begin{table}[ht]
\caption{Resource update characteristics for each action mode.}
\centering
\begin{tabular}{ccc}
\specialrule{.1em}{.05em}{.05em}
 Mode & Power & Data \\
\specialrule{.05em}{.05em}{.05em}
$i$ & $\dot{p}_\ell < 0$ & $\dot{d}_\ell > 0$ \\
$g$ & $\dot{p}_\ell < 0$ & $\dot{d}_\ell < 0$ \\
$\mathsc{nil}$ & $\dot{p}_\ell \ge 0$ &  $\dot{d}_\ell > 0$\\
\specialrule{.1em}{.05em}{.05em}
\end{tabular}
\label{tab:resource_states}
\end{table}

\subsection{Reward Function}

The reward function $R(s,a)$ is the associated reward for taking action $a$ in state $s$. The reward function is designed to encode mission-specific limitations or desired behavior. The specific choice of function comes from expert-knowledge of the mission designers and engineers. In this work, we define reward to be initally $0$ and add to it for each of the conditions listed in \Cref{tab:reward_states} that are satisfied by the input state-action pair $(s,a)$.

\begin{table}[ht]
\caption{Initial position error for different orbit determination techniques}
\centering
\begin{tabular}{cc}
\specialrule{.1em}{.05em}{.05em}
 Reward & Condition \\
\specialrule{.05em}{.05em}{.05em}
$\gamma^{(t_s - t)}r_i$ & if $\ell \in I$ and $\ell \notin I_c$ \\
$1\times 10^{-1} \times (t_s - t)$ & if $\ell = \mathsc{g}$ \\
$1\times 10^{-4} \times (t_s - t)$ & if $\ell = \mathsc{nil}$ \\ 
$-1\times 10^4$ & $p \leq p_{\text{min}}$ \\
$-1\times 10^4$ & $d \geq d_{\text{max}}$ \\
\specialrule{.1em}{.05em}{.05em}
\end{tabular}
\label{tab:reward_states}
\end{table}

The first condition is the reward for collecting image $i$ if that location has not been previously imaged. The imaging action has a discount factor $\gamma$ applied to properly discount the reward for actions further into in the planning horizon relative to actions closer in time. If omitted, the reward for an action at the very end of the planning horizon would be the same as an action at the start. The second condition provides a reward for downlinking data proportional to the length of the contact. The third condition encourages taking sun-pointed actions by assigning a small reward for the duration of the action. The final two conditions apply a strong penalty if the spacecraft violates safe power or data thresholds.

\subsection{Implementation Challenges}

Even with a semi-Markov model, we could not solve the problem exactly. The state space grows exponentially with just the number of images being scheduled. This neglects the size of state space dimensions in time, previous action, and the discretization of power and data. Because the size of the space grows at least exponentially, finding an exact solution quickly becomes computationally intractable even for small sets of image collection requests. For this reason, we do not consider exact MDP solutions algorithms, such as value iteration or policy iteration, and instead use approximate, online solution methods.

Using online solutions alone was not enough to enable the problem to be solved quickly given the size of the possible action space. Consider the size of the action space $A(s)$ for the initial state. For a problem of $1000$ locations, there are typically $2000$ to $4000$ image collection opportunities in a \SI{24}{\hour} planning horizon, almost all of which would be feasible actions from the initial state. Most of these opportunities are detrimental to take as they skip earlier collection opportunities that could have otherwise been taken. For these reasons, we impose a limit on the number of actions in $A(s)$ to the first $N_a^{\text{max}}$ in time.

Finally, when using an online solution method, the action space is a function of the state and must be recomputed for each state visited, computed a-priori, or cached in some manner. We found that recomputing the action space at each state works for the forward-search solution, but the MCTS approach relies too heavily on simulation rollouts to make recomputation practical. To make MCTS computationally feasible we precompute the action space.

\subsection{Solution Methods}

To solve the problem, we used common algorithms for solving MDPs and SMDPs\textemdash forward search and Monte Carlo Tree Search. While both are discussed in greater depth in other works \cite{sutton2018reinforcement,kochenderfer2015decision,browne2012survey}, we will present a brief review of the techniques here for understanding.

\subsubsection{Forward Search}

Forward search is an online method that exhaustively explores the state and action space from the current state $s$ to some depth $d_{\text{solve}}$.
Shown in \Cref{alg:forward_search}, it uses \textsc{SelectAction} to iterate over all possible actions in the current state, recursively calling itself on future states until the desired depth is reached. 

Solving the SMDP proceeds from the initial state $s_0$ by computing all paths of depth $d_{\text{solve}}$, then choosing the action associated with the path of highest reward. The transition function is called to deterministically advance the state. In the next state, the process is repeated, and the algorithm may find a path different than the previous maximizes the reward. This sequence is repeated until the end of the planning period $H$. The resulting series of states and actions is the task plan.

The solve depth $d_{\text{solve}}$ varies based on computational resources and the desired optimality of the solution. The choice of $d_{\text{solve}}$ is also inherently tied to the choice of the actions space size limit $N_a^{\text{max}}$. This can be seen by the total computational complexity of this method, which is $O(|$$N_a^{\text{max}}|^{d_{\text{solve}}})$. Larger values of $d_{\text{solve}}$ will lead to better plans due to greater ``foresight'' of the algorithm when choosing actions but come at increased computational cost.

\begin{algorithm}[htb]
\caption{SMDP Forward Search}
\label{alg:forward_search}
\begin{algorithmic}[1]
\Function{SelectAction}{$s,d_{\text{solve}},\gamma$}
\If{$d = 0$}
	\State \Return (\textsc{nil}, 0)
\EndIf
	\State $(a^\star, v^\star) \gets (\texttt{NIL}, -\infty)$
\For{$a \in A(s)$}
	\State $v \gets R(s,a)$
	\For{$s^\prime \in T(s, a)$}
		\State $(a^\prime, v^\prime) \gets \textsc{SelectAction}(s^\prime, d_{\text{solve}}-1,\gamma )$
		\State $v \gets v + \gamma^{(a.t_s - s.t)}v^\prime$
	\EndFor
	\If{$v > v^\star$}
		\State $(a^\star, v^\star) \gets (a, v)$
	\EndIf
\EndFor
\State \Return $(a^\star, v^\star)$
\EndFunction
\Function{ForwardSearch}{$s_0,d_{\text{solve}},\gamma$}
\State $\pi(s_0) \gets \left[\;\right]$
\State $a, v \gets \textsc{SelectAction}(s_0, d_{\text{solve}},\gamma )$
\State $s \gets T(s_0,a)$
\State $\pi(s_0) \gets \textsc{Append}(\left[(s_0, a)\right])$
\While{$v \neq -\infty$}
	\State $a, v \gets \textsc{SelectAction}(s, d_{\text{solve}},\gamma )$
	\State $s \gets T(s,a)$
	\State $\pi(s_0) \gets \textsc{Append}(\left[(s, a)\right])$
\EndWhile
\State \Return $\pi(s_0)$
\EndFunction
\end{algorithmic}
\end{algorithm}

\subsubsection{Monte Carlo Tree Search}

Monte Carlo Tree Search is an online, sampling-based approach \cite{browne2012survey,kocsis2006bandit,gelly2011monte}. In contrast to forward search, the algorithm's computational complexity does not grow exponentially with the solve depth or size of the action space. The approach entails running many simulations from the current state, updating an estimate of the state-action value function on each iteration. After a fixed number of simulations have been run, the best action found is taken, and the process repeats until the end of the planning horizon is reached.

\Cref{alg:mcts} presents the method. In the algorithm, $Q(s,a)$ is the estimate of state-action value and $N(s,a)$ is the number of times that the pair has been observed during simulation. The function $Q_0(s,a)$ and $N_0(s,a)$ allow the algorithm to be seeded if prior knowledge is available. In this work, they are both set to 0. The hyperparameter $c$ controls the amount of exploration during the search, and $N_{\text{sim}}^{\text{max}}$ sets is maximum number of simulations performed for each solution step. Exploration done at one step helps inform subsequent steps because $Q(s,a)$ persists between steps.

\begin{algorithm}[htb]
\caption{SMDP Monte Carlo Tree Search}
\label{alg:mcts}
\begin{algorithmic}[1]
\Function{Simulate}{$s,d_{\text{solve}},\gamma$}
\If{$d = 0$}
	\State \Return 0
\EndIf
\If{$s \notin V$}
	\For{$a \in A(s)$}
		\State $N(s,a),Q(s,a) \gets N_0(s,a),Q_0(s,a)$
	\EndFor
	\State $V \gets V \cup \left\{s\right\}$
	\State \Return $\textsc{Rollout}(s, d_{\text{solve}},\gamma)$
\EndIf
\State $a \gets \text{argmax}_a Q(s,a) + c\sqrt{\frac{\log \sum_{a \in A(s)}N(s, a)}{N(s,a)}}$
\State $r \gets R(s, a)$
\State $s^\prime \gets T(s, a)$
\State $q \gets \gamma^{(a.t_s - s.t)}\textsc{Simulate}(s^\prime, d_{\text{solve}}-1,\gamma)$
\State $N(s,a) \gets N(s,a) + 1$
\State $Q(s,a) \gets Q(s,a) + \frac{q-Q(s,a)}{N(s,a)}$
\State \Return $q$
\EndFunction
\Function{Rollout}{$s,d_{\text{solve}},\gamma$}
\If{$d = 0$}
	\State \Return 0
\EndIf
\State $a \sim A(s)$
\State $r \gets R(s,a)$
\State $s^\prime \gets T(s, a)$
\State \Return $r + \gamma^{(a.t_s - s.t)}\textsc{Rollout}(s^\prime, d_{\text{solve}}-1,\gamma)$
\EndFunction
\Function{MonteCarloTreeSearch}{$s_0,d_{\text{solve}},\gamma$}
\State $\pi(s_0) \gets \left[\;\right]$
\State $n \gets 0$
\For{$n \leq N_{\text{sim}}^{\text{max}}$}
	\State $\textsc{Simulate}(s_0,d_{\text{solve}},\gamma)$
	\State $n \gets n + 1$
\EndFor
\State $a \gets \text{argmax}_a Q(s,a)$
\State $s \gets T(s_0,a)$
\State $\pi(s_0) \gets \pi(s_0) \cup \left\{(s_0, a)\right\}$
\While{$v \neq -\infty$}
	\State $n \gets 0$
	\For{$n \leq N_{\text{sim}}^{\text{max}}$}
		\State $\textsc{Simulate}(s,d_{\text{solve}},\gamma)$
		\State $n \gets n + 1$
	\EndFor
	\State $a \gets \text{argmax}_a Q(s,a)$
	\State $s \gets T(s,a)$
	\State $\pi(s_0) \gets \textsc{Append}(\left[(s, a)\right])$
\EndWhile
\State \Return $\pi(s_0)$
\EndFunction
\end{algorithmic}
\end{algorithm}

\section{Baselines}

We compare our methods against serval approaches used previously to solve the satellite task scheduling problem. For this work, we consider three of the most common approaches to provide a performance baseline for comparison\textemdash a rule-based method, a graph search approach, and a MILP formulation.

\subsection{Rule-Based Planning}

The rule-based method is simply hand-coded logic based on expert knowledge of the system and desired behavior. We use an adaptation of a prior implementation \cite{bianchessi2008planning}. Shown in \Cref{alg:rule_solution}, the rule-based planner is simple in nature. It looks at the next feasible action in time. If this feasible action is a collect or contact, and there are enough resources available, it will take this action. Otherwise, it will sun-point for an equivalent duration. The algorithm terminates when no action can be taken at the current state.

\begin{algorithm}[thb]
\caption{Rule-Based Task Planner}
\label{alg:rule_solution}
\begin{algorithmic}[1] 
\Function{NaiveSelectAction}{$s$}
\State // Let $a$ be the next action in time from $A(s)$
\If{$a \in G \cup I$}
	\State $s^\prime \gets T(s,a)$
	\If{$s^\prime.p > p_{min}$ and $s^\prime.d \leq d_{max}$}
		\State // Take action if feasible given resources
		\State \Return $a$
	\Else
		\State // Sunpoint for equivalent duration
		\State \Return $a_{sunpoint}(t_s)$
	\EndIf
\Else
	\State \Return $a_{sunpoint}(t_s)$
\EndIf
\EndFunction
\Function{RuleBasedPlanner}{$s_0,d_{\text{solve}},\gamma$}
\State $\pi(s_0) \gets \left[\;\right]$
\State $a \gets \textsc{NaiveSelectAction}(s_0)$
\State $s \gets T(s_0,a)$
\State $\pi(s_0) \gets \pi(s_0) \cup \left\{(s_0, a)\right\}$
\While{$A(s) \neq \emptyset$}
	\State $a \gets \textsc{NaiveSelectAction}(s)$
	\State $s \gets T(s,a)$
	\State $\pi(s_0) \gets \textsc{Append}(\left[(s, a)\right])$
\EndWhile
\State \Return $\pi(s_0)$
\EndFunction
\end{algorithmic}
\end{algorithm}

\subsection{Graph Search}

The graph search method works by first encoding the action space and feasible transitions as a directed acyclic graph then computing the longest weighted path that traverses the graph. In this approach, each possible collect opportunity is a node and feasible transitions between collects are edges. The optimal plan is then simply the longest weighted path through the graph, where the weights are the rewards for collecting each image. 

\Cref{alg:graph_solution} allows for the direct calculation of this path through dynamic programming. Here, $R_i$ is the total reward collected for the plan found passing through node $i$, and $n_i$ is the prior node associated with the highest reward. We first initialize all opportunities $o_i$ collecting image $j$ to the reward function $r_j$ associated to collecting that image. The optimal path is found by iterating through nodes in increasing time order, updating the highest value path through each node, then back-tracking from the highest value node found with $\text{argmax}_i R_i$. The agility constraint function $C(t_s^p, t_s)$ is extended to return feasibility of the transition between collection opportunities $C(o_1,o_2)$. If $C(o_1,o_2) = 0$, then it is not possible to take both opportunities $o_1$ and $o_2$. The tasking plan is the reverse of all nodes (collection opportunities) traversed while back-tracking. This approach is based on the work of Augenstein \cite{augenstein2014optimal}.

\begin{algorithm}[htb]
\caption{Graph-based Task Planner}
\begin{algorithmic}[1] 
\Function{PropagateWeights}{}
	\For{$o_i \in O_j, j \in I$}
		\State $R_i \gets r_j$
		\State $n_i \gets \mathsc{nil}$
	\EndFor
	\For{$o_i \in O_j, j \in I$}
		\For{$o_k \in O_\ell, \ell \in I$}
			\If{$C(o_i,o_k) = 1 \And R_i + r_\ell > R_k$}
				\State $R_k \gets R_i + r_\ell$
				\State $n_k \gets i$
			\EndIf
		\EndFor
	\EndFor
\EndFunction
\Function{ExtractPath}{}
	\State $\pi(s_0) \gets \left[\;\right]$
	\State $n \gets \text{argmax}_x R_x$
	\While{$n \neq \mathsc{nil}$}
		\State $\pi(s_0) \gets \textsc{append}\left(\left[o_n\right]\right)$
		\State $n \gets p_n$
	\EndWhile
	\State $\pi(s_0) \gets \textsc{Reverse}(\pi(s_0))$
	\State \Return $\pi(s_0)$
\EndFunction
\end{algorithmic}
\label{alg:graph_solution}
\end{algorithm}

\subsection{Mixed-Integer Linear Programming}

The third reference approach is a mixed-integer linear programming (MILP) formulation based on the work of Nag et al. \cite{nag2018scheduling}. This approach attempts to solve
\begin{equation*}
\begin{aligned}
{\text{maximize}} \hspace{0.5cm} & \sum_{x_{ij}} r_ix_{ij} \; \forall o_j \in O_i, i \in I \\
\text{subject to} \hspace{0.5cm} & x_{ij} \in \{0, 1\} \\
& \sum_{o_j \in O_i} x_{ij} \leq 1 \; \forall i \in I \\
& x_{ij} + x_{k\ell} \leq 1 \; \forall o_j \in O_i, i \in I \\
& \hspace{2.075cm} \forall o_\ell \in O_k, k \in I \\
& \hspace{2.075cm} \text{s.t.} \; C(o_j,o_\ell) = 0 \\
\end{aligned}
\label{eqn:mlip_solution}
\end{equation*}
Here, the problem is formulated in terms of in terms of binary variable $x_{ij}$ which indicates whether the image $i$ is collected at opportunity $o_j$ ($x_{ij}=1$) or not ($x_{ij}=0$). The objective represents the total reward from all collect opportunities across all requested images and stations. The first constraint is the integer constraint on the binary decision of whether the collect is taken or not. The second constraint imposes a limit of at most one collect taken per each image to ensure that the solution attempts to capture all images. The final constraint enforces that all possible transitions between collection opportunities must obey constraints on spacecraft agility. The task schedule is the set of collection opportunities selected.

\section{Experiments}

To evaluate the performance of the SMDP planning solution, we create  hypothetical sets of imaging locations that are randomly drawn the the LandSat-2 imaging catalog. The Landsat program regularly collects images of the entire globe along the pre-defined WRS-2 grid. The data set contains 16,896 coastal or land image locations. All locations are assigned an equal reward value of $r_i = 1$. 

We consider two variations of the satellite tasking problem. First, we consider the performance in the more common resource-free problem addressed by the baseline approaches. We also consider the formulation where spacecraft resources of power and data are included. In the later case, the rule-based approach is used to provide the comparison as the graph and MILP formulations cannot easily extended to incorporate resources into the planning process.

All scenarios consider a single spacecraft in a \SI{500}{\kilo\meter} circular polar orbit. We assume that the spacecraft has a \SI{1}{\degree\per\second} maximum slew rate between pointing-vectors. The spacecraft model has a 10\% duty cycle on power, one image collection uses 1\% of total data storage capacity. Spacecraft also generates data at constant rate of 0.0001\% the maximum capacity every second to model on-board telemetry logging. Data can be downlinked at 4 times the rate at which it is generated during image collection. We set $p_{\text{min}}$ to 30\% of the total energy capacity and $d_{\text{max}}$ to 75\% of total data storage. The spacecraft dynamics are propagated using the SGP4 propagator \cite{vallado2006revisiting}. The planning horizon is set to be 1 day long for all cases. For all planning simulations, the image collection, ground contact opportunities, and action space were precomputed and reused for all solutions. All simulations were run on a workstation with dual 14-core 2.6 GHz Intel E5-2690 processors with each core running at most one simulation at a time.

\begin{figure*}[tb!]
\centering
\includegraphics[height=3in]{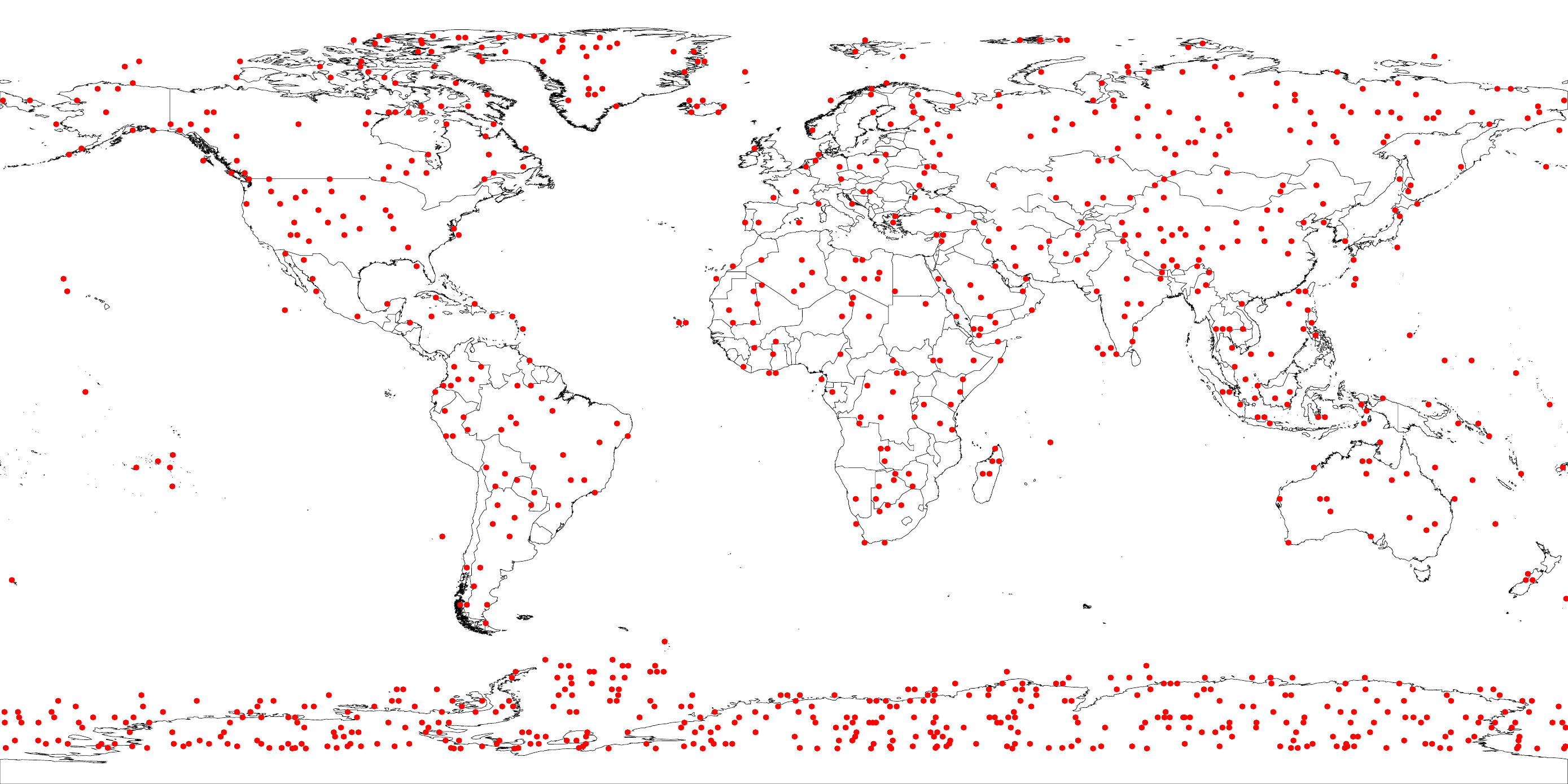}
\caption{Random sampling of 1000 locations from the Landsat grid data set. This specific sampling was used to conduct hyperparameter searches.}
\label{fig:landsat_locations}
\end{figure*}

\subsection{Hyperparameter Selection}

Both SMDP solution algorithms rely on a number of hyperparameters that affect their performance (in terms of speed and optimality). Before comparing the SMDP algorithms, it is important to understand the parameters used to control the algorithms. To discover the best possible set of parameters, we performed a brute-force grid search of a select few parameters to find the best combination to use for the resource-free and resource-modeled problems. Grid search was used to find the best parameter combination. The same set of 1000 locations shown in \Cref{fig:landsat_locations} was used for all searches.

\begin{table}[ht]
\centering
\begin{tabular}{cc}
\specialrule{.1em}{.05em}{.05em}
 Parameter & Considered values \\
\specialrule{.05em}{.05em}{.05em}
$\gamma$ & $0.99, 0.9925, 0.995, 0.999, 0.9995, 0.9999$ \\ \vspace{1pt}
$d_{\text{solve}}$ & $3, 5, 7$ \\ \vspace{1pt}
$N_a^{\text{max}}$ & $3, 4, 5$ \\
\specialrule{.1em}{.05em}{.05em}
\end{tabular}
\caption{Hyperparameter search space for forward search.}
\label{tab:hp_search_fs}
\end{table}

\begin{table}[ht]
\centering
\begin{tabular}{cc}
\specialrule{.1em}{.05em}{.05em}
 Parameter & Considered values \\
\specialrule{.05em}{.05em}{.05em}
$\gamma$ & $0.99, 0.995, 0.999$ \\ \vspace{1pt}
$d_{\text{solve}}$ & $5, 10, 15, 20$ \\ \vspace{1pt}
$c$ & $0.1, 1.0, 3.0, 5.0, 10.0$ \\ \vspace{1pt}
$N_{\text{sim}}^{\text{max}}$ & $50, 100, 200, 500$ \\ \vspace{1pt}
$N_a^{\text{max}}$ & $3, 4, 5$ \\
\specialrule{.1em}{.05em}{.05em}
\end{tabular}
\caption{Hyperparameter search space for MCTS. Ten simulations were performed per point in the parameter space.}
\label{tab:hp_search_mcts}
\end{table}

\begin{table}[ht]
\centering
\begin{tabular}{cccc}
\specialrule{.1em}{.05em}{.05em}
$N_a^{\text{max}}$ & $\gamma$ & $d_{\text{solve}}$ & Reward \\
\specialrule{.05em}{.05em}{.05em}
\multicolumn{1}{l}{\textit{w/o Resources}} & \phantom{-} & \phantom{-} & \phantom{-} \\
3 & 0.999 & 3 & 450 \\
4 & 0.999 & 3 & 449 \\
5 & 0.999 & 3 & 447 \\
\multicolumn{1}{l}{\textit{w/ Resources}} & \phantom{-} & \phantom{-} & \phantom{-} \\
3 & 0.99 & 7 & 225 \\
4 & 0.9925 & 3 & 223 \\
5 & 0.995 & 3 & 221 \\
\specialrule{.1em}{.05em}{.05em}
\end{tabular}
\caption{Hyperparameter outcomes for forward search.} 
\label{tab:hp_search_fs_results}
\end{table}

\begin{table}[ht]
\centering
\begin{tabular}{cccccc}
\specialrule{.1em}{.05em}{.05em}
$N_a^{\text{max}}$ & $\gamma$ & $d_{\text{solve}}$ & $c$ & $N_{\text{sim}}^{\text{max}}$ & Reward \\
\specialrule{.05em}{.05em}{.05em}
\multicolumn{1}{l}{\textit{w/ Resources}} \\
3 & 0.995 & 10 & 3.0 & 500 & 420 \\
4 & 0.995 & 10 & 5.0 & 500 & 415 \\
5 & 0.995 & 10 & 3.0 & 500 & 406 \\
\multicolumn{1}{l}{\textit{w/o Resources}} \\
3 & 0.99 & 20 & 3.0 & 200 & 215 \\
4 & 0.99 & 20 & 3.0 & 50 & 204 \\
5 & 0.999 & 15 & 3.0 & 100 & 195 \\
\specialrule{.1em}{.05em}{.05em}
\end{tabular}
\caption{Hyperparameter outcomes for MCTS. Reward is the mean value over 10 simulations.}
\label{tab:hp_search_mcts_results}
\end{table}

\Cref{tab:hp_search_fs_results,tab:hp_search_mcts_results} show   the hyperparameter combination with the highest mean reward for each value of $N_a^{\text{max}}$ considered. It is interesting that the planning reward does not increase as the limit on the action space size increases. Instead, the reward decreases slightly. In the case of forward search, this is attributed to the fact that the larger decision space adds actions that are further out in time. These actions, while possibly better in the short term, ultimately advance the solution more quickly in time and reduces the total number of requests that can be collected.

In the case of MCTS, there are two explanations for the decreasing reward with action space size. When resource management is not part of the problem, the larger action space means that more iterations are required to achieve similar levels of exploration across the decision tree. The maximum number of simulations is capped at 500. Consequently, as the action space size increases, the algorithms are unable to explore as thoroughly and performs worse. 

When resource management is part the problem formulation, having a greater look-ahead dominates the MCTS planning outcomes as evidenced by greater $d_{\text{solve}}$ values. While one would expect that more rollout simulations should lead to better planning outcomes in MCTS, it is likely that the increased solve depth causes enough statistical variance that 10 iterations per parameter combination are not enough to find the optimal hyperparameter combination.

\subsection{Planning without Resources}

\begin{figure*}
\begin{center}
\includegraphics[width=\textwidth]{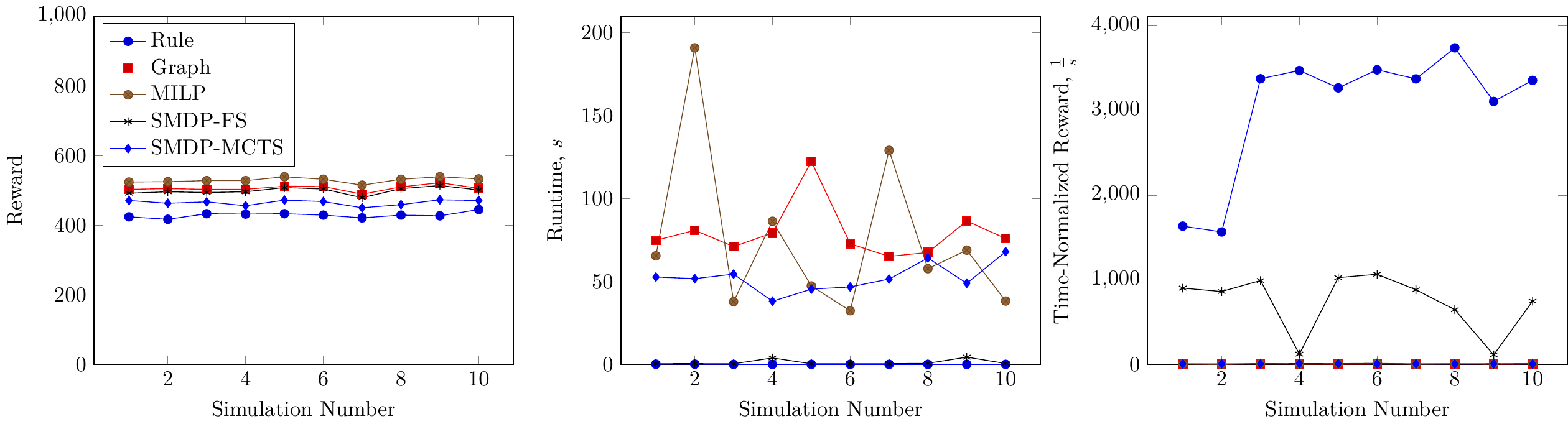}
\end{center}
\caption{Performance of planning approaches without considering resources for 10 random samplings of 1000 locations of from Landsat data. Total reward (left), simulation runtime (middle), and time-normalized reward (right). The time-normalized reward is the simulation reward divided by the runtime.}
\label{fig:smdp_results_no_resources_1000}
\end{figure*}

\Cref{fig:smdp_results_no_resources_1000} shows how rule-based, graph, MILP, SMDP forward search, and SMDP MCTS formulations perform when only the single objective of maximizing overall reward is considered. As expected, the rule-based method is the least performant but most computationally efficient. The MILP approach consistently performs the best out of all methods but is also one of the slowest. The next best method is the graph search approach. SMDP forward search manages to be nearly as efficient as the graph formulation, while having computational efficiency near that of the rule-based approach. It also significantly outperforms the reward of the rule-based approach in all cases. The comparative efficiency of forward search is hightlighted in the right panel of \Cref{fig:smdp_results_no_resources_1000}, where it has the best efficiency rating behind the rule-based method for reward per second of computation time. Even though the action space is precomputed, the MCTS solution is still computationally expensive relative to the rule-based and forward search solutions due to the number of iterations required to thoroughly sample the state-action space.

\subsection{Planning with Resources}

\begin{figure*}
\begin{center}
\includegraphics[width=\textwidth]{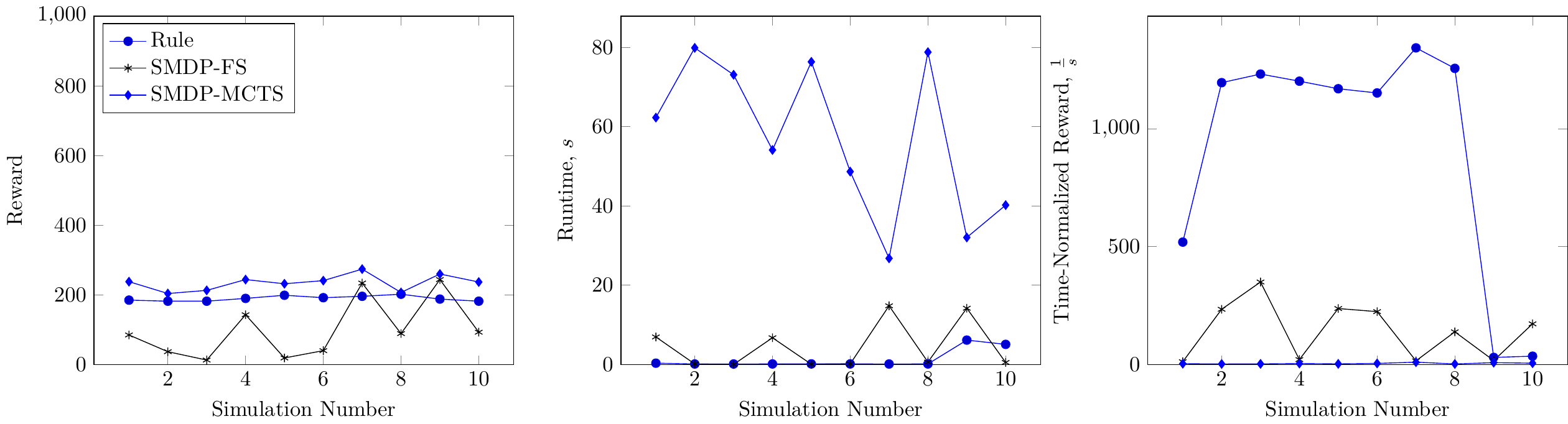}
\end{center}
\caption{Performance of planning approaches considering resource management for 10 random samplings of 1000 locations of from Landsat data. Total reward (left), simulation runtime (middle), and time-normalized reward (right). The time-normalized reward is the simulation reward divided by the runtime.}
\label{fig:smdp_results_yes_resources_1000}
\end{figure*}

When resources are considered as part of the scheduling problem, task planning becomes significantly more challenging. The system must balance multiple objectives of imaging, downlinking data, and recharging to maximize the overall reward. \Cref{fig:smdp_results_yes_resources_1000} shows that out of the three solution methods considered the SMDP MCTS approach is the best at handling this challenge and forward search the worst. While it might be surprising that a rule-based method can outperform a more complex approach like forward search, it is important to remember that with a short solve depth the algorithm becomes myopic\textemdash paths that initially look promising can end up steering the solution towards ultimately inefficient outcomes. It is expected that this could be alleviated by increasing the solve depth. However, due to the exponential complexity of the forward search algorithm in solution depth, increasing the solve depth is not a computationally practical solution. Monte Carlo Tree Search can support significantly greater solve depths. Consistent with the hypothesis that solve depth is the driving factor behind making efficient trades between multiple objectives, MCTS performs best with resource management. The rule-based method, while only having an effective solve depth of a single step, ends up being better on-average than forward search. 

We ran simulations for 200, 500, and 2000 location sets both with and without considering resource management. Plots from these simulations are included as part of the Appendix.

\section{Conclusions}

This paper introduced a formulation of the satellite task scheduling problem as a semi-Markov Decision process. Two viable solution methods for the problem, forward search and Monte Carlo Tree Search, were provided. Both techniques are shown capable of generating task optimized task schedules. Forward search is shown to be a simple and computationally efficient approach that is most effective in planning when only considering a single objective. MCTS is shown to be able to effectively balance multiple objectives while planning and captures the most reward out of the methods considered. In future work, it would desirable to extend the MILP formulation to also include resources to provide an additional baseline for comparison. Additionally, the SMDP approaches would benefit from a more thorough investigation of hyperparameters and their effect on scheduling outcomes. It would also be interesting to consider, higher fidelity resource models and transition functions, as well as extend the action space to include maneuvering.

\appendices{}              

\begin{figure*}
\begin{center}
\includegraphics[width=\textwidth]{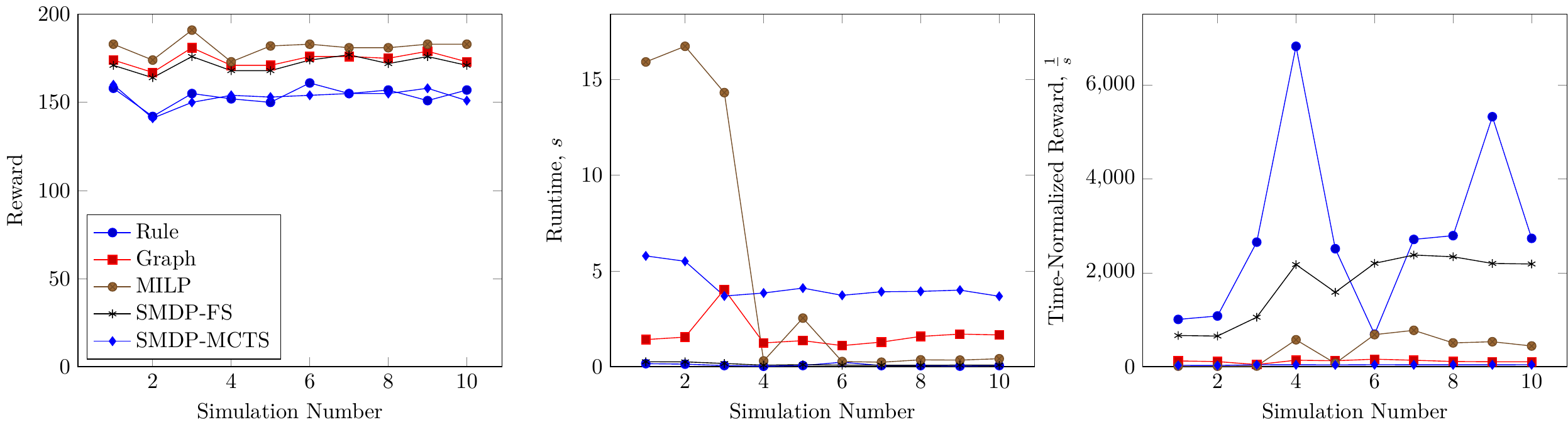}
\end{center}
\caption{Performance of planning approaches without considering resources for 10 random samplings of 200 locations of from Landsat data. Total reward (left), simulation runtime (middle), and time-normalized reward (right). The time-normalized reward is the simulation reward divided by the runtime.}
\label{fig:smdp_results_no_resources_200}
\end{figure*}

\begin{figure*}
\begin{center}
\includegraphics[width=\textwidth]{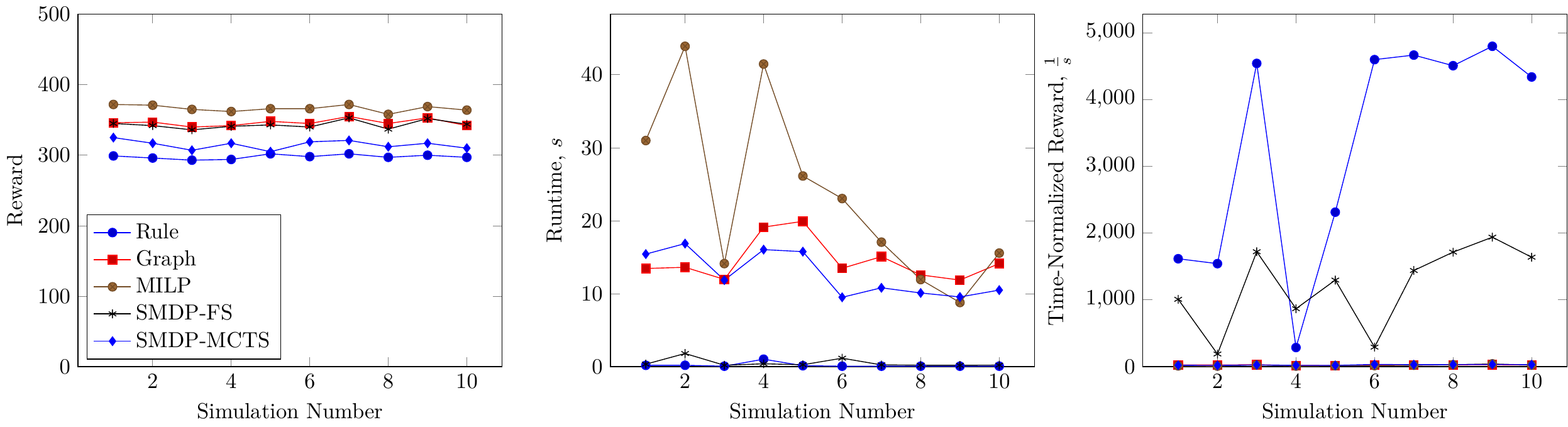}
\end{center}
\caption{Performance of planning approaches without considering resources for 10 random samplings of 500 locations of from Landsat data. Total reward (left), simulation runtime (middle), and time-normalized reward (right). The time-normalized reward is the simulation reward divided by the runtime.}
\label{fig:smdp_results_no_resources_500}
\end{figure*}

\begin{figure*}
\begin{center}
\includegraphics[width=\textwidth]{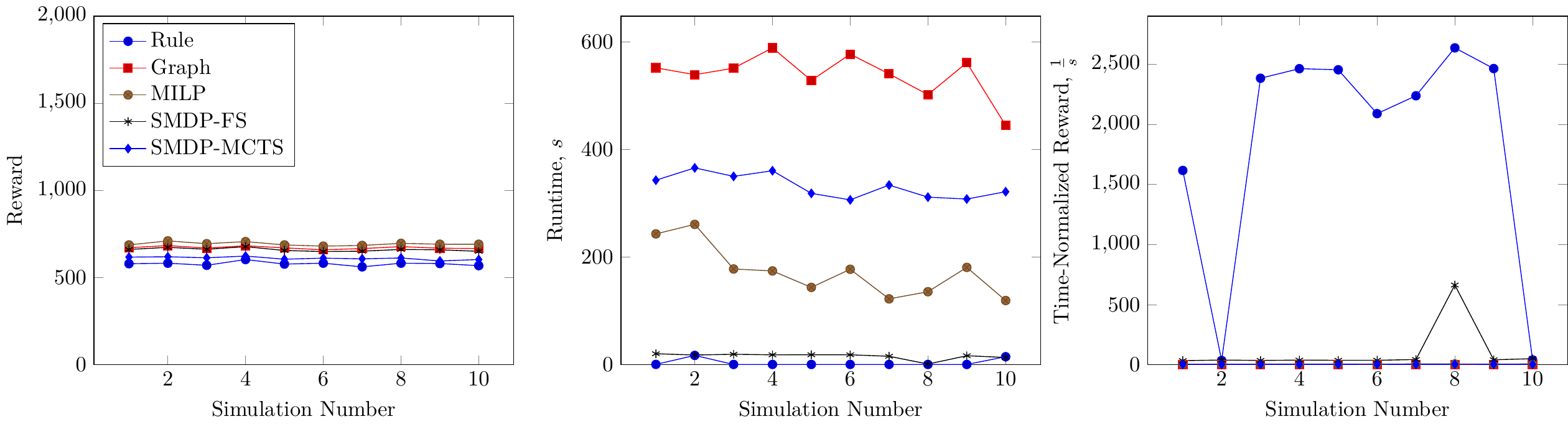}
\end{center}
\caption{Performance of planning approaches without considering resources for 10 random samplings of 2000 locations of from Landsat data. Total reward (left), simulation runtime (middle), and time-normalized reward (right). The time-normalized reward is the simulation reward divided by the runtime.}
\label{fig:smdp_results_no_resources_2000}
\end{figure*}

\begin{figure*}
\begin{center}
\includegraphics[width=\textwidth]{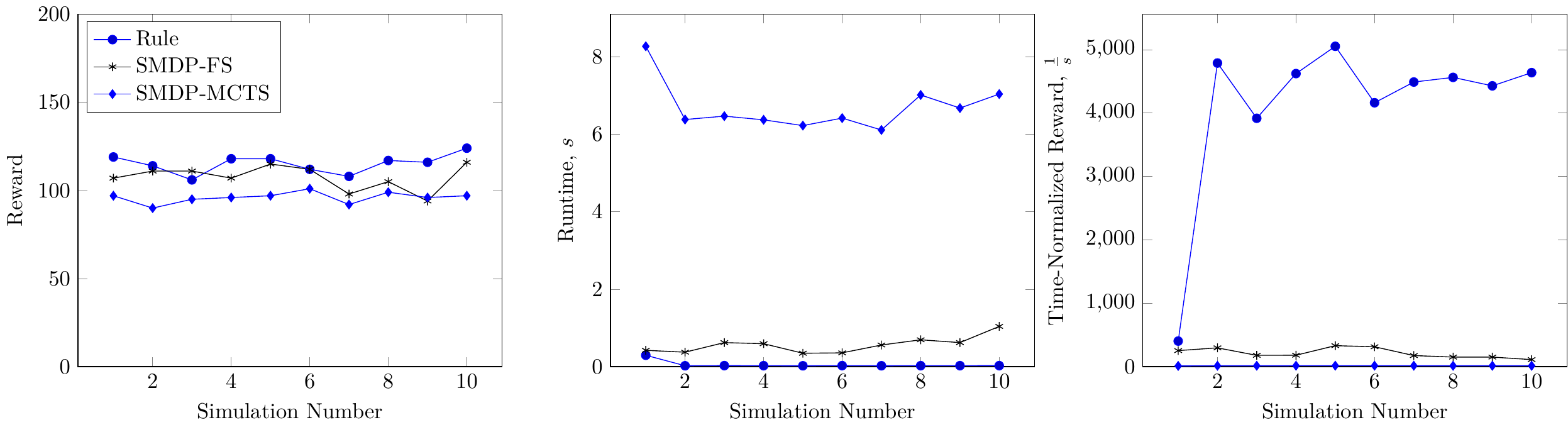}
\end{center}
\caption{Performance of planning approaches considering resource management for 10 random samplings of 200 locations of from Landsat data. Total reward (left), simulation runtime (middle), and time-normalized reward (right). The time-normalized reward is the simulation reward divided by the runtime.}
\label{fig:smdp_results_yes_resources_200}
\end{figure*}

\begin{figure*}
\begin{center}
\includegraphics[width=\textwidth]{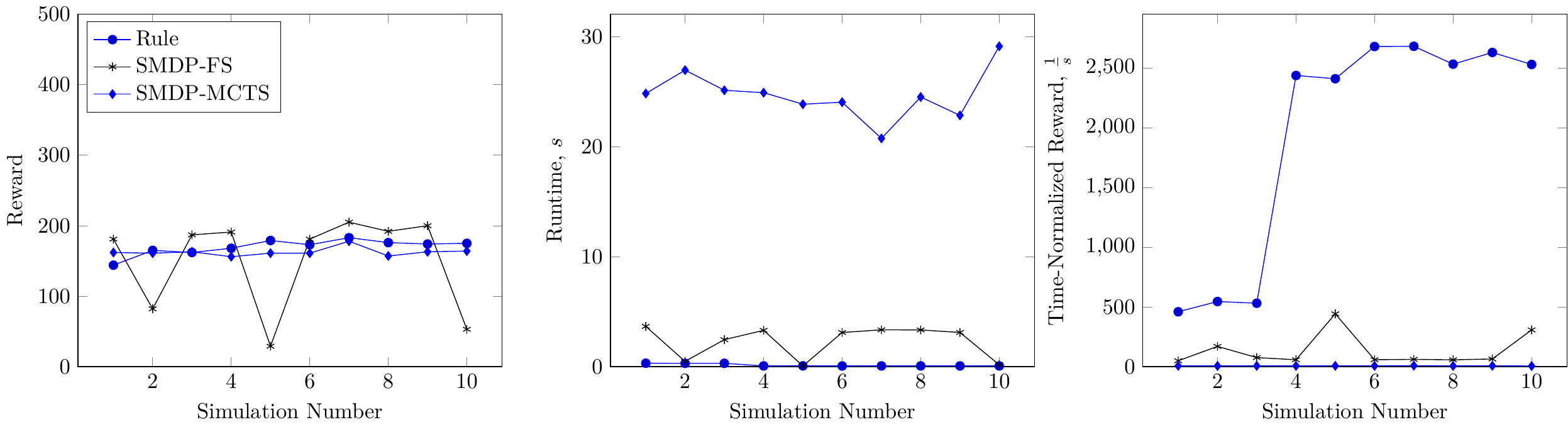}
\end{center}
\caption{Performance of planning approaches considering resource management for 10 random samplings of 500 locations of from Landsat data. Total reward (left), simulation runtime (middle), and time-normalized reward (right). The time-normalized reward is the simulation reward divided by the runtime.}
\label{fig:smdp_results_yes_resources_500}
\end{figure*}

\begin{figure*}
\begin{center}
\includegraphics[width=\textwidth]{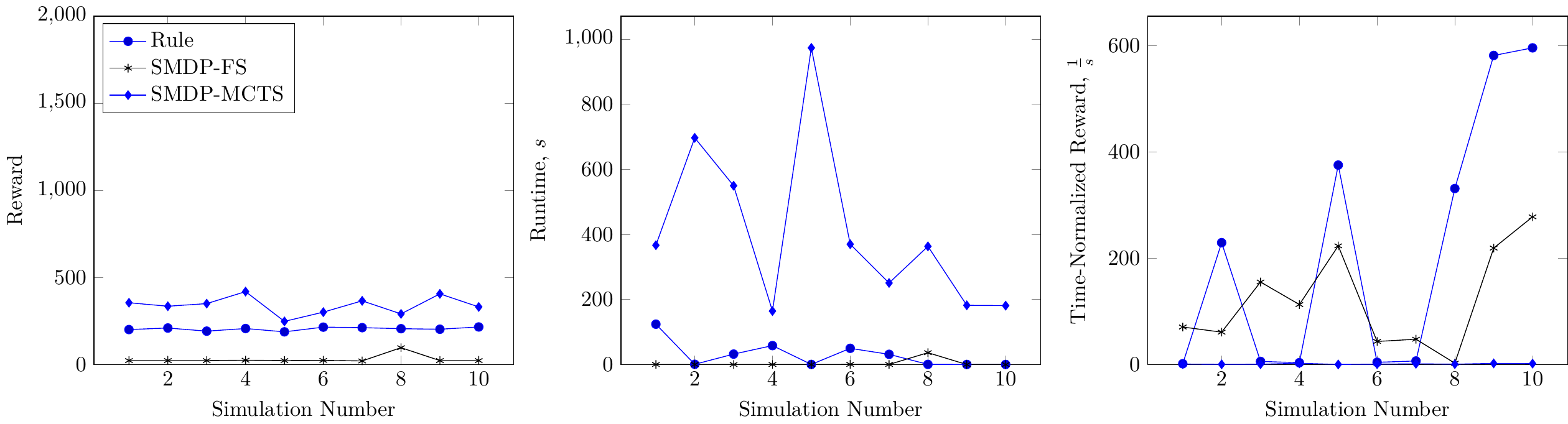}
\end{center}
\caption{Performance of planning approaches considering resource management for 10 random samplings of 2000 locations of from Landsat data. Total reward (left), simulation runtime (middle), and time-normalized reward (right). The time-normalized reward is the simulation reward divided by the runtime.}
\label{fig:smdp_results_yes_resources_2000}
\end{figure*}

To evaluate performance of the SMDP forward search and Monte Carlo Tree Search algorithms, a number of simulations were run beyond 1000-location simulations presented as part of the results section. \Cref{fig:smdp_results_no_resources_200,fig:smdp_results_no_resources_500,fig:smdp_results_no_resources_2000} provide the single-objective results for 200, 500, and 2000 location sets.  \Cref{fig:smdp_results_yes_resources_200,fig:smdp_results_yes_resources_500,fig:smdp_results_yes_resources_2000} provide the multi-objective results for 200, 500, and 2000 location sets.

Interestingly, when considering planning with multiple objectives, forward search proves to be more efficient than the rule-based method when the problem size is small (100), but quickly becomes the least effective as the problem size grows. As with the general shortcomings of forward search in multi-objective optimization, it is thought the short solve depth of the forward search method is unable to detect and make efficient trades between the multiple objectives.

\acknowledgments
The authors would like to acknowledge Lucas Riggi for providing the initial algorithm for computing imaging opportunities, Sreeja Nag for providing the Landsat image location data set, Christian Lenz for his continual support and encouragement in undertaking this research. We would also like to thank Amir Maleki for his insightful feedback while preparing this publication.

\bibliographystyle{IEEEtran}
\bibliography{deddy_research.bib}

\thebiography
\begin{biographywithpic}
{Duncan Eddy}{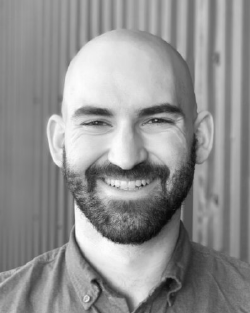}
is a PhD candidate in Stanford Intelligent Systems Laboratory (SISL) in the department of Aeronautics and Astronautics at Stanford University. His research focuses on autonomous spacecraft operations and decision making in the space domain. He received B.S. in Mechanical Engineer from Rice University in 2013, and M.S. in Aerospace Engineering from Stanford University in 2015. He is also the Director of Space Operations at Capella Space Corporation. In this capacity he oversaw the successful launch of the first commercial radar satellite in the United States, and continues to work at Capella developing highly automated mission operations systems. 

\end{biographywithpic} 

\begin{biographywithpic}
{Mykel Kochenderfer}{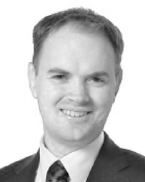}
 is Assistant Professor of Aeronautics and Astronautics and Assistant Professor, by courtesy, of Computer Science at Stanford University. He is the director of the Stanford Intelligent Systems Laboratory (SISL), conducting research on advanced algorithms and analytical methods for the design of robust decision making systems. Prior to joining the faculty in 2013, he was at MIT Lincoln Laboratory where he worked on airspace modeling and aircraft collision avoidance. He received his Ph.D. from the University of Edinburgh in 2006 where he studied at the Institute of Perception, Action and Behaviour in the School of Informatics. He received B.S. and M.S. degrees in computer science from Stanford University in 2003.

\end{biographywithpic}

\end{document}